# Wakefield generation and electron acceleration via propagation of radially polarized laser pulses in homogeneous plasma

Shivani Aggarwal, Saumya Singh, Dinkar Mishra, Bhupesh Kumar[1] and Pallavi Jha[2]

*Department of Physics, University of Lucknow, Lucknow (U.P.)– 226007, India*

## Abstract

The paper presents a study of wakefield generation and electron injection via propagation of radially polarized laser pulses in homogeneous pre-ionized plasma. The analytical study is based on Lorentz force and continuity equations. Perturbation technique and quasi-static approximation are used for evaluating the generated longitudinal wakefields. Trapping and acceleration of electrons are examined by injecting a test electron in the generated wakefields. The results are compared with those obtained via linearly polarized laser pulses. The validation of analytical results is performed using the Fourier-Bessel particle-in-cell (FBPIC) simulation code. It is seen that there is a significant enhancement in amplitude of the longitudinal wakefield generated and electron energy gain via radially polarized laser pulses as compared to linearly polarized laser pulse case.

---

[1] Corresponding Author: bhupeshk05@gmail.com

[2] Retired Professor

# 1. Introduction

Interaction of intense laser pulses with homogeneous or inhomogeneous plasma gives rise to several nonlinear effects such as self-focusing and self-channeling [1-3]. Laser-plasma interactions play a crucial role in many important applications like inertial confinement fusion [4-5], laser-wakefield acceleration [6], x-ray lasers [7-8], high-power laser-field and ionized plasma interaction [9-12]. The pioneering work on laser-plasma interaction by Tajima and Dawson in 1979 demonstrated that an intense laser pulse propagating through plasma creates ponderomotive force, responsible for generating an electron-plasma wave called wakefield [13-14]. These wakefields are used to accelerate self-injected as well as externally injected electrons in the laser wakefield accelerator (LWFA) [15].

In recent years, there has been growing interest in generating high-power radially polarized laser beams due to their advantages in various fields of applications. In a radially polarized laser, the polarization vector is always directed towards the center of the pulse with presence of strong longitudinal electric field at the focus. Various techniques like superposition of lasers and optical elements such as gratings, and plasma mirrors have been used to generate tightly focused radially polarized laser pulses [16-17]. Analytical and simulation studies have been done to study the fields and propagation characteristics of radially polarized laser pulses in vacuum and it is noticed that these fields show less diffraction in comparison to other polarization states [18]. Due to tight focusing capability, they are being used in high-resolution microscopy, material processing, optical trapping, metal cutting, and direct laser acceleration [17-21]. It is noticed that a radially polarized beam accelerates electrons and keeps them tightly focused near the optic axis via direct laser acceleration process [22]. Studies have shown that in comparison with linearly polarized and circularly polarized laser pulses, radially polarized laser pulses are more useful for wakefield generation due to their focusing characteristics and beam quality. Hence, they are expected to lead to higher electron energy gain [23-27].

Analytical studies have shown that enhanced energy gain can be obtained using a periodic frequency chirped radially polarized laser pulse propagating in vacuum as compared to linearly chirped pulse [23]. Powell et al have shown that radially polarized laser beam leads to enhancement in the trapping and acceleration of an electron in vacuum so that by the inherent complete symmetry of radially polarized laser beam an electron can be accelerated to the level of GeV [24]. Simulation studies have shown that radially polarized laser pulses produce better beam quality with less energy spread as compared to linearly polarized laser pulses. Axial component of the laser field can be used to accelerate proton bunches initially focused by radial component of the laser field in vacuum [28-30]. Multi particle simulation studies have shown that radially polarized laser pulses can be used for GeV range acceleration in vacuum for axially injected particles like electrons and alpha particles without much diffraction [31]. Marceau et al have shown that electron bunches generated using a low-density gas target can be accelerated axially with low energy spread using a non-paraxial, ultrafast radially polarized laser pulse [26]. Simulation studies have shown that a radially polarized laser pulse gives electron acceleration with low energy spread in a long plasma micro-channel having parabolic density in the transverse radial direction [12].

This paper deals with an analytical study of wakefield generation and electron injection mechanism via propagation of radially polarized laser pulses in homogeneous plasma. The use of radially polarized laser pulses for wakefield generation and particle acceleration may be preferred over other polarized laser pulses as it remains focused on axis and can be propagated for longer distances with lesser divergence [32]. The paper is organized as follows: In section 2, mathematical formulation and longitudinal electric wakefield generation, graphical findings and comparison with linearly polarized laser pulses are discussed. In section 3, Simulation studies and the validation of analytically obtained results are presented. Section 4, discusses test electron trapping and acceleration of injected electrons in the longitudinal wakefields generated behind the laser pulse. Summary and conclusions are presented in section 5.

## 2. Mathematical Formulation

Consider a radially polarized laser pulse propagating paraxially along the z-direction in homogeneous plasma having ambient plasma density $n_o$. The electric $(\vec{E})$ and magnetic $(\vec{B})$ components of the radially polarized laser pulse [30] can be written in cylindrical coordinates as

$$\vec{E} = E_0 exp\left(\frac{-r^2}{r_0^2}\right) g(z,t) \left(\hat{r}\left(\frac{r}{2r_0}\right) \cos(k_o z - \omega_o t) + \hat{z}\left(1 - \frac{r^2}{r_0^2}\right) \sin(k_o z - \omega_o t)\right), \qquad (1)$$

$$\vec{B} = \hat{\theta} E_0 \frac{r}{2r_0} exp\left(\frac{-r^2}{r_0^2}\right) g(z,t) \cos(k_o z - \omega_o t), \qquad (2)$$

where $E_0$ is the amplitude of the laser field, $g(z,t) = exp(-(z-ct)^2/L^2)$ represents the spatial envelope of the laser pulse and $c, L, \lambda_o, k_o, r_o, \omega_o$ are respectively speed of light in vacuum, pulse length, wavelength, wave number, radius at beam waist and frequency of the radially polarized pulse. The motion of plasma electrons in the presence of these fields, in homogeneous plasma, is governed by Lorentz force and continuity equations given by,

$$\frac{d\vec{v}}{dt} = \frac{\partial \vec{v}}{\partial t} + (\vec{v}.\vec{\nabla})\vec{v} = -\frac{e}{m}\left[\vec{E} + \frac{1}{c}(\vec{v} \times \vec{B})\right], \qquad (3)$$

and

$$\frac{\partial n}{\partial t} + \vec{\nabla}.(n\vec{v}) = 0, \qquad (4)$$

where $n$ represents the perturbed plasma density and $\vec{v}$ is the plasma electron velocity. In order to obtain slow electric wakefields lowest order, slow density perturbations are evaluated. This requires evaluation of perturbed plasma electron velocity.

Considering perturbative expansion of Eq. (3) for moderate laser intensity, the first order quiver velocity of plasma electron is given by,

$$\vec{v}^{(1)} = -a_o c \left[ -\hat{r} A_1 \sin(k_o z - \omega_o t) + \hat{z} \left( 1 + \frac{\omega_p^2}{\omega_o^2} \right) A_2 \cos(k_o z - \omega_o t) \right], \tag{5}$$

where $A_1 = \frac{r}{2r_0} exp\left(\frac{-r^2}{r_0^2}\right) g(z,t)$, $A_2 = \left(1 - \frac{r^2}{r_0^2}\right) exp\left(\frac{-r^2}{r_0^2}\right) g(z,t)$ and $a_o = \frac{eE_o}{mc\omega_o}(\ll 1)$ is the laser strength parameter. Plasma electrons interacting with radially polarized laser pulses have quiver velocity with components along r and z directions. From Eq. (5), it may be noted that the radial component of plasma electron velocity vanishes at $r = 0$ but for $r \neq 0$, transverse components should also exist in present case. Therefore, in case of a very tightly focused pulse only axial plasma electron velocity exists. Using Eq. (5), the first order perturbative expansion of Eq. (4) gives the perturbed fast oscillating density as,

$$n^{(1)} = -n_o a_o A_2 \cos(k_o z - \omega_o t), \tag{6}$$

This first order density perturbation arises on account of the z component of quiver velocity and is zero for a linearly polarized laser field. With the help of Eq. (5), the second order perturbative expansion of Eq. (3) gives the equation governing the slow plasma electron velocity as,

$$\frac{\partial \vec{v}^{(2)}}{\partial t} = -\frac{e}{m} \vec{E}_w^{(2)} - \frac{(a_o c)^2}{4} \vec{\nabla}(A_1^2 + A_2^2), \tag{7}$$

where the generated, slow electric wakefield driven by second order density perturbation is represented by Gauss' law $\frac{\partial}{\partial z} E_w^{(2)} = -4\pi e n^{(2)}$.

The second order perturbative expansion of Eq. (4) gives,

$$\frac{\partial n^{(2)}}{\partial t} + \vec{\nabla} \cdot \left(n^{(0)} \vec{v}^{(2)}\right) = 0. \tag{8}$$

It may be noted that the first order density perturbation coupled with the quiver velocities contributes significantly to the second order density perturbation. In order to obtain longitudinal wakefield $E_{zw}^{(2)}$, Eq. (8) is transformed in terms of independent variables $\xi = z - ct$ and $\tau = t$. Further, quasi-static approximation (QSA) is applied to the transformed equation. Under QSA, it is assumed that the laser pulse amplitude does not evolve significantly as it transits a plasma electron. Eliminating $n^{(2)}$ and substituting Gauss' law, yields the transformed equation describing the evolution of the longitudinal electric wakefield as,

$$\left(\frac{\partial^2}{\partial \xi^2} + k_p^2\right) E_{zw}^{(2)} = -\frac{mc^2 k_p}{4e}\left(\frac{\partial}{\partial \xi}(A_1^2 + A_2^2)\right) \tag{9}$$

where $k_p (= \omega_p/c)$ is the wave number of the plasma wave and $\omega_p = (4\pi n_o e^2/m)^{1/2}$ is the plasma frequency. Solving Eq. (9) gives,

$$E_{zw}^{(2)} = \varepsilon \, \frac{k_p^2 L}{4} \sqrt{\frac{\pi}{2}} e^{-(k_p^2 L^2/8)} \left[\left(\frac{r}{2r_0}\right)^2 + \left(1 - \frac{r^2}{r_0^2}\right)^2\right] \cos(k_p \xi). \tag{10}$$

where $E_{zw}^{(2)}$ represents the longitudinal electric wakefield generated behind the laser pulse and $\varepsilon = \left(\frac{mc^2 a_0^2}{e} e^{-(2r^2/r_0^2)}\right)$. Since the laser pulse profile is Gaussian, maxima of the generated wakefield (Eq. (10)) is obtained for $\lambda_p = L\pi\sqrt{2}$, and is given by

$$E^{(2)}_{zwm} = E_A \cos(k_p \xi) \tag{11}$$

where $E_A = \varepsilon \frac{k_p^2 L}{4} \sqrt{\frac{\pi}{2}} e^{-(k_p^2 L^2/8)} \left[\left(\frac{r}{2r_0}\right)^2 + \left(1 - \frac{r^2}{r_0^2}\right)^2\right]$ represents the amplitude of the maximized longitudinal wakefield.

In order to study the variation of maximized longitudinal electric wakefield with transverse distance (r), the amplitude of the wakefield ($E_A$) is plotted as shown in Fig. 1 for parameters $a_o = 0.3$, $\lambda_o = 0.8\ \mu m$, $r_o = 15\ \mu m$, $L = 12\ \mu m$, $n_o = 3.8 \times 10^{17} cm^{-3}$.

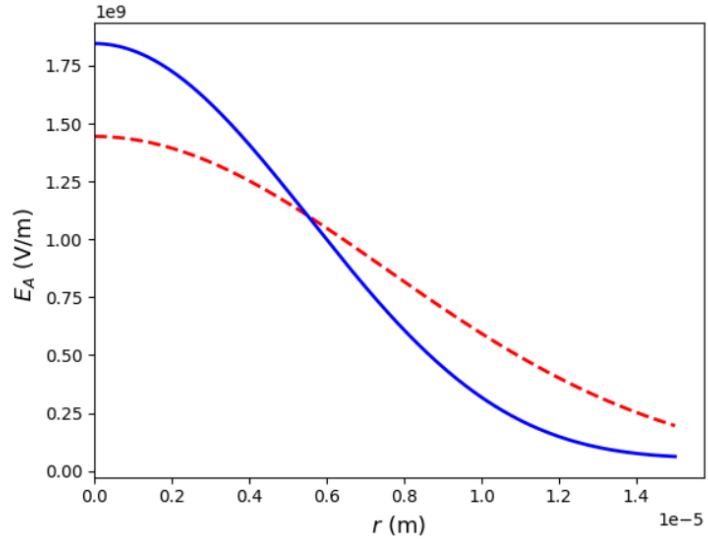

Fig. 1. Variation of amplitude of the maximized longitudinal wakefield $E_A$ showing solid curve for radially polarized laser pulse and dashed curve for linearly polarized laser pulse with respect to transverse distance $r$ for $a_o = 0.3$, $\lambda_o = 0.8\ \mu m$, $r_o = 15\ \mu m$, $L = 12\ \mu m$, $n_o = 3.8 \times 10^{17} cm^{-3}$.

The solid curve represents the amplitude of the maximized longitudinal electric wakefield generated by radially polarized Gaussian laser pulse with respect to the transverse distance $r$ while the dashed curve shows the amplitude of the maximized longitudinal electric wakefield generated by a linearly polarized Gaussian laser pulse. It is seen that the amplitude of the maximized longitudinal wakefield is maximum on-axis (r = 0) for both cases. On account of the complex dependence on r, the amplitude decreases with r for radially polarized laser pulses while an exponential decrease is observed for the linearly polarized pulse. It is observed that both curves intersect at $r = 5.65\ \mu m$. Due to rapid decrease in wake amplitude for the radially polarized laser, the wakefield generated by the linearly polarized laser is seen to be larger beyond the intersection point. However, the amplitudes for both cases are significantly reduced, as compared to the near axis fields.

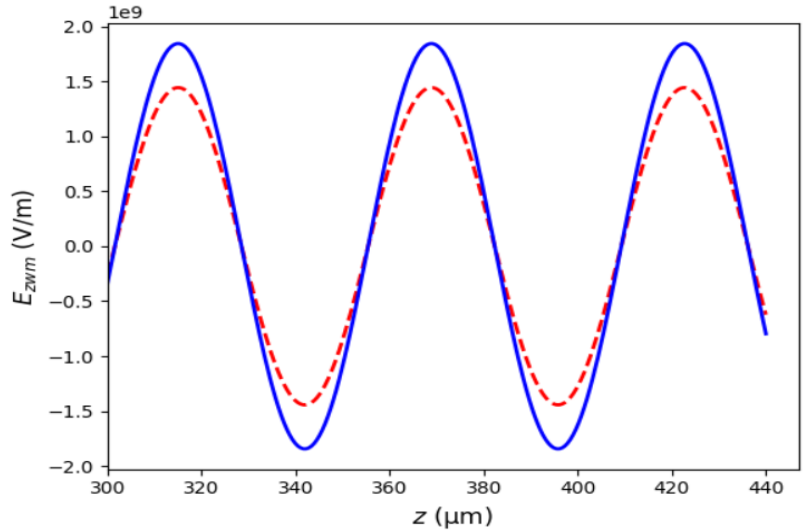

Fig. 2. Variation of amplitude of the maximized longitudinal electric wakefield $E_{zwm}$ at on axis generated by radially polarized laser pulse (solid curve) and linearly polarized laser pulse (dashed curve) with respect to propagation distance $z$ for $a_o = 0.3,\ \lambda_o = 0.8\ \mu m,\ r_o = 15\ \mu m,\ L = 12\ \mu m,\ n_o = 3.8 \times 10^{17} cm^{-3}$.

Figure 2 represents the sinusoidal variations of on-axis longitudinal electric wakefield $E_{zwm}$ with respect to the propagation distance $z$ for $a_o = 0.3,\ \lambda_o = 0.8\ \mu m,\ r_o = 15\ \mu m,\ L = 12\ \mu m,$ $n_o = 3.8 \times 10^{17} cm^{-3}$. The solid curve represents the longitudinal electric wakefield generated by radially polarized Gaussian laser pulse with respect to the propagation distance $z$ while the dashed curve shows the longitudinal electric wakefield generated by a linearly polarized Gaussian laser pulse. It is seen that amplitude of the wakefield generated by radially polarized ($1.8 \times 10^9$ V/m) is 28.5% higher than that obtained via linearly polarized ($1.4 \times 10^9 V/m$) pulses. The enhanced amplitude may be attributed to the presence of strong axial component of the laser electric field. Consequently, electrons experience an additional force along the axial direction.

## 3. Simulation

In order to validate the analytical results Fourier-Bessel Particle-In-Cell (FBPIC) simulations are performed for laser and plasma parameters used for the analytical study. The radially polarized laser pulse having laser spot size ($r_0$) equal to 15 $\mu m$ and pulse length 12 $\mu m$ is launched from the left side of the simulation box of size $130 \mu m$ (longitudinal direction) $\times\ 25\ \mu m$ (transverse direction). The number of grid points along the axial direction is 1200 and in the transverse direction is 150. The simulation setup utilizes an electromagnetic PIC grounded in the Fourier-Bessel decomposition. Similar simulation run is performed for the linearly polarized laser pulse.

Figure 3 depicts that amplitude of the wakefield generated by radially polarized ($1.6 \times 10^9\ V/m$) is 16.67% higher than that obtained via linearly polarized ($1.2 \times 10^9 V/m$) laser pulses. while the analytical analysis shows 28.5% enhancement in amplitude of the wakefield generated by radially polarized laser pulse with respect to linearly polarized laser pulse. It is observed that simulation results are comparable to the analytical results obtained. However, in both the cases, simulation results of wakefield amplitudes are slightly lower than that obtained in the analytical calculations. This difference may be attributed to Gouy phase consideration in simulations which is not considered in calculating wakefields analytically.

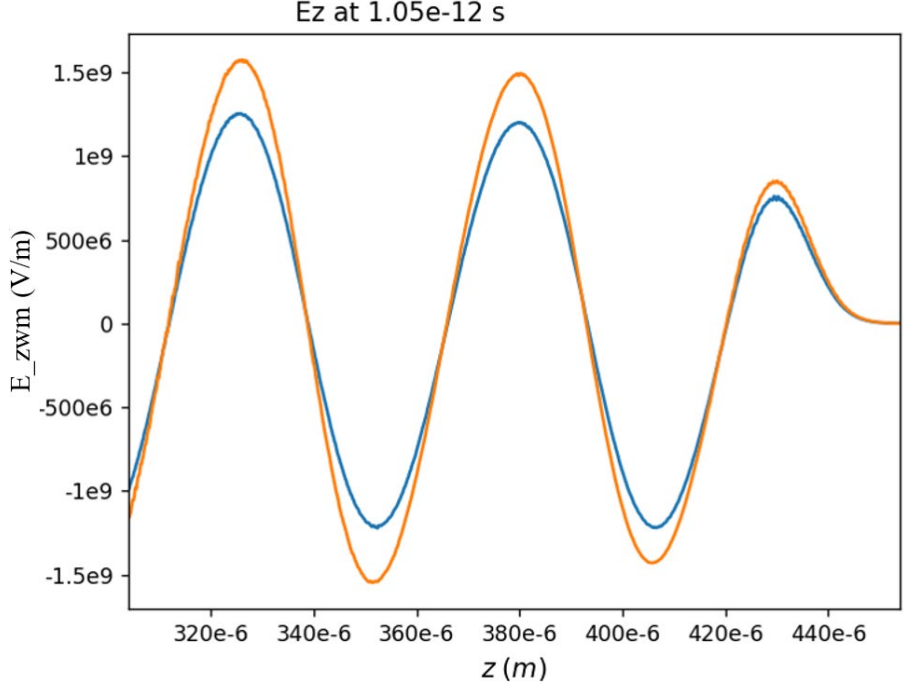

Fig. 3. Simulation results for wakefield generation via radially polarized laser pulses (red line) and linearly polarized laser pulses (blue line) for $a_o = 0.3$, $\lambda_o = 0.8\ \mu m$, $r_o = 15\ \mu m$, $L = 12\ \mu m$, $n_o = 3.8 \times 10^{17} cm^{-3}$, $\omega_p = 0.35 \times 10^{14}\ s^{-1}$.

**4. Electron acceleration**

In order to study electron acceleration and energy gain, consider a test electron injected along the z direction, in the generated wakefield. The force (longitudinal component) experienced by the electron is given by,

$$F_{zm} = \frac{d(m\gamma_e v_e)}{dt} = -eE_{zwm}, \quad (12)$$

where $\gamma_e = (1 - v_e^2/c^2)^{-1/2}$ and $v_e$ are the relativistic factor and velocity of the test electron respectively. Substituting Eq. (11) into Eq. (12) gives the rate of change of injected electron energy as,

$$\frac{d\gamma_e}{dt} = -\frac{e}{mc}\left(1 - \frac{1}{\gamma_e^2}\right)^{1/2} \varepsilon \, \frac{k_p^2 L}{4} \sqrt{\frac{\pi}{2}} e^{-(k_p^2 L^2/8)} \left[\left(\frac{r}{2r_0}\right)^2 + \left(1 - \frac{r^2}{r_0^2}\right)^2\right] \cos(\Psi). \tag{13}$$

where $\Psi(= k_p \xi)$ is the relative phase of the test electron with respect to the wakefield. Dividing Eq. (13) by $(d\Psi/dt)$ and integrating with respect to $\Psi$ gives,

$$\gamma_e - \beta_p(\gamma_e^2 - 1)^{1/2} = \varepsilon \, \frac{k_p^2 L}{4} \sqrt{\frac{\pi}{2}} e^{-(k_p^2 L^2/8)} \left[\left(\frac{r}{2r_0}\right)^2 + \left(1 - \frac{r^2}{r_0^2}\right)^2\right] \sin(\Psi) + C_s, \tag{14}$$

where $\beta_p = \omega_p/ck_p$ and $C_s = (\gamma_{eo} - \beta_p(\gamma_{eo}^2 - 1)^{1/2})$ is the constant of integration. Eq. (14) is used to depict the phase space trajectory of the test electron in the wakefield generated by propagation of radially polarized laser pulses in homogeneous plasma.

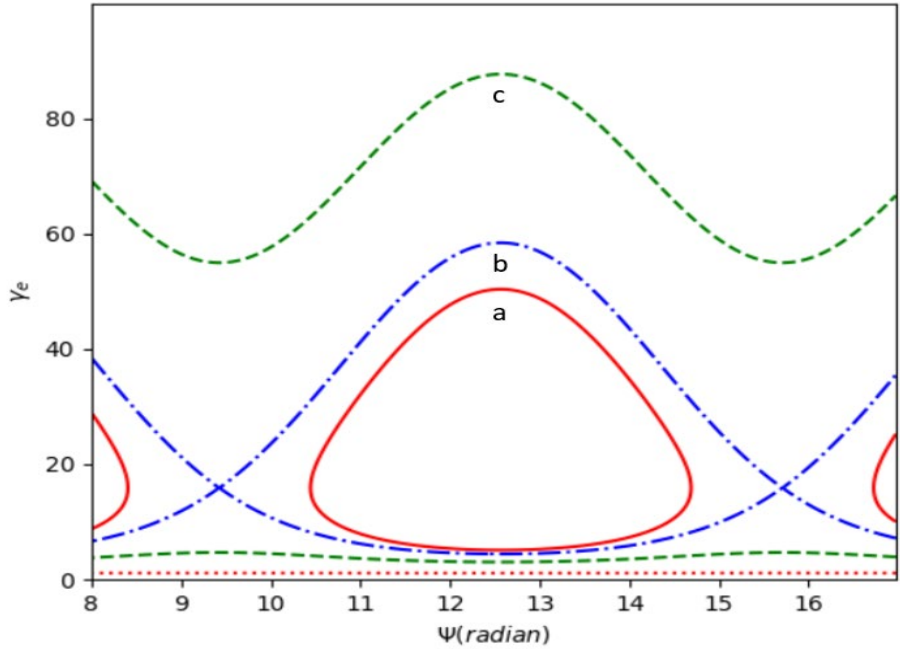

Fig. 4. Phase space plot for radially polarized laser pulses showing (a) trapped, (b) separatrix, (c) untrapped orbits for initial energy (a) 2.55 MeV, (b) 1.53 MeV and (c) 1.02 MeV injected for $a_o = 0.3, \lambda_o = 0.8\ \mu m, r_o = 15\ \mu m, L = 12\ \mu m, n_o = 3.8 \times 10^{17} cm^{-3}, \omega_p = 0.35 \times 10^{14}\ s^{-1}$.

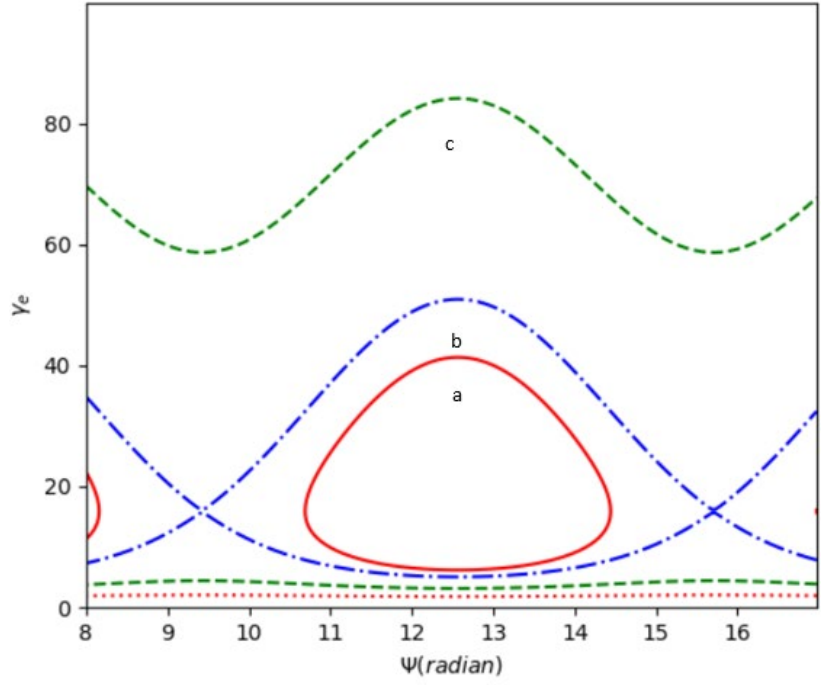

Fig. 5. Phase space plot for linearly polarized laser pulses showing (a) trapped, (b) separatrix, (c) untrapped orbits for initial energy (a) 3.06 MeV, (b) 2.55 MeV and (c) 2.04 MeV injected for $a_o = 0.3, \lambda_o = 0.8\ \mu m, r_o = 15\ \mu m, L = 12\ \mu m, n_o = 3.8 \times 10^{17} cm^{-3}, \omega_p = 0.35 \times 10^{14}\ s^{-1}$.

Test electron phase space trajectories for radially polarized and linearly polarized laser pulses, are plotted in Figs. 4 and 5 respectively, for the same parameters as used for Fig. 1. Curves (a), (b) and (c) represent the trapped test electron trajectory, the separatrix that separates the trapped and untrapped electron orbits and untrapped electron orbits respectively. Electrons riding the separatrix obtained via radially (linearly) polarized laser pulses require a minimum injection energy of 1.53 MeV (2.04 MeV). The maximum energy attained $[(\gamma_{emax} - 1)mc^2]$ is 29.63 MeV (25.03 MeV). Hence the gain in energy $[(\gamma_{emax} - \gamma_{eo})mc^2]$ is 28.10 MeV (22.48 MeV). Hence radially polarized laser pulses lead to higher electron energy using lower injection energy and therefore significantly higher gain, than that obtained via the linearly polarized laser pulses.

## 5. Summary and conclusions

In this paper, an analytical study of wakefield generation due to interaction of radially polarized laser pulse with homogeneous plasma have been studied. Lorentz force and continuity equations are used to derive the fast as well as slow oscillating velocities and density perturbation. Quasi-static approximation (QSA) is applied to the transformed density equation. QSA is used to obtain slow density perturbation and wakefields. Longitudinal electric wakefields generated behind the radially polarized laser pulses have been calculated and compared with wakefields generated via linearly polarized laser pulses. It is observed that wakefield generated by radially polarized laser pulses ($1.8 \times 10^9\ V/m$) is 28.5% higher than that obtained via linearly polarized laser pulses ($1.4 \times 10^9 V/m$). The analytical results have been validated via simulation studies. Further, phase space analysis shows that a test electron injected externally into the generated wakefield with appropriate phase can be trapped and accelerated to higher energies using radially polarized laser pulses. Energy gain of a test electron injected into wakefields generated by radially polarized laser pulse is nearly 28% higher than the gain obtained via linearly polarized laser pulse.

**Data availability statement**

Data that supports the findings of the present study are contained within the study.

**Author Declarations**

**Conflict of Interest**

The authors have no conflicts to disclose.